\documentclass[aps,amsmath,amssymb,reprint]{revtex4-1}	
\usepackage{graphicx}
\usepackage{dcolumn}
\usepackage{bm}
\usepackage{amsmath,amsthm,amssymb}

\begin{document}
\title{Aspects of $C_3$-symmetric calorons from numerical Nahm transform}

\author{Atsushi Nakamula$^a$}
\email{nakamula@sci.kitasato-u.ac.jp}
\author{Nobuyuki Sawado$^b$}
\email{sawado@ph.noda.tus.ac.jp}
\author{Koki Takesue$^b$}
\email{j6213614@ed.tus.ac.jp}
\affiliation{$^a$Department of Physics, School of Science, Kitasato University,
Sagamihara, Kanagawa 252-0373, Japan,\\
$^b$Department of Physics, Tokyo University of Science,
 Noda, Chiba 278-8510, Japan}


\date{\today}

\begin{abstract}
Calorons are finite action solutions to the anti-selfdual Yang-Mills equations on $\mathbb{R}^3\times S^1$.
They are generally constructed by the so called Nahm construction. 
We perform the numerical Nahm transform for the Nahm data of 3-calorons with $C_3$-symmetry, which
do not have the monopole limits.
Dissimilar to the cases of having monopole limits, we can trace the zero-circumference limit of $S^1$.
It is found that the action density of the calorons tends to fade away as  $S^1$ shrinks.
\end{abstract}
\pacs{Valid PACS appear here}

\maketitle

\section{Introduction}
Topological solitons in various field theories are vigorously studied in both analytically and numerically \cite{ManSut}.
In this paper, we consider the finite action solutions to anti-selfdual Yang-Mills gauge theory in $\mathbb{R}^3\times S^1$, usually called calorons \cite{HS}.
As a topological soliton, calorons have connection with many other related objects.
In particular, they may have a non-trivial holonomy around the $S^1$-direction, which links them to Skyrmions \cite{Zahed}.
And also, there exists a picture of calorons in which monopoles constitute them \cite{AH,Shnir}.
Finally, when the circumference of $S^1$ becomes large, then calorons will turn into instantons on $\mathbb{R}^4$.
Hence the studies on calorons will give insight into the universal understanding of topological solitons.
The topological indices of calorons are thus holonomy, monopole charges and instanton charges.
Among them, the holonomy and instanton charges  do not take  integer valued in general \cite{Etesi_Jardim_2008}.

It is well known that calorons can be produced generally through the Nahm construction \cite{Nahm_1984}, 
in which  the dual space description of the gauge fields, called the Nahm data, plays central role. 
In the construction, the transformation from the description on the dual space into that on the ordinary configuration space is crucial,
and called the Nahm transform. The Nahm transform for $SU(2)$ caloron of instanton charge $k=1$ with non-trivial holonomy was studied analytically in
\cite{KvB, Lee:1998bb}. For the higher charges, exact calorons of $k=2$ were found in \cite{BNvB}. 

There are many studies for the numerical Nahm transform, {\it e.g.}, \cite{GonzalezArroyo:1998cq,GarciaPerez:1999bc}.
Recently, the 2-caloron Nahm data of monopole charge $(m_1,m_2)=(2,2)$ of arbitrary mass with 16 moduli parameters has been 
proposed \cite{NakaSaka}. The numerical Nahm transform for the 10 parameter subset of the data 
has been done in \cite{Muranaka_Nakamula_Sawado_Toda_2011}. 
In this paper we employ the scheme developed in \cite{Muranaka_Nakamula_Sawado_Toda_2011}, but
we make an essential refinement of it to be applicable for general cases.

As mentioned above, calorons are closely related to monopoles and instantons.
In fact, the Nahm data of calorons have usually the monopole, or the large scale, limits in addition to the instanton limits.
In other words, most of the known Nahm data give examples of the picture in which calorons interpolate between monopoles and instantons \cite{Harland}.
However, it is known that the calorons from Corrigan-Fairlie-t'Hooft ansatz \cite{CF}, \textit{i.e.}, the Harrington-Shepard type,  
of higher instanton charges do not have the monopole limits \cite{Ward_2004}.
In addition, it has been shown that there is a one-parameter family of 3-calorons Nahm data with $C_3$-symmetry which do not have monopole
 limits \cite{Nakamula_Sawado_2012,Muranaka_Nakamula_Sawado_2013}.
From these facts, it will be interesting to ask that under which conditions the calorons have monopole limits.
In this context, we consider the numerical Nahm transform for the $C_3$-symmetric 3-calorons and show the visualization of their action density.
Unlike the cases of having monopole limits, we can trace the zero-circumference limit of $S^1$, \textit{i.e.}, 
the limit to $\mathbb{R}^3$.
We find that the action density gradually tends to fade away as $S^1$ becomes smaller.

In this paper, we restrict the gauge group to $SU(2)$ or $U(2)$.
In section 2, a brief review on the Nahm construction is given for the case that the holonomy is trivial.
In section 3, we give the method of numerical Nahm transform.
In section 4, the Nahm data of the $C_3$-symmetric 3-calorons are introduced.
In section 5, the visualization for the action density is performed for the caloron Nahm data given in the previous section.
The final section is devoted for summary and concluding remarks.

\section{Nahm construction}
We give a review of the Nahm construction for the calorons of $U(2)$ gauge theory with trivial holonomy.
Let $x^{\mu}$ be the standard coordinates on $\mathbb{R}^{3}\times S^1$, 
where the greek indices run $1,2,3$ and $4$.

Caloron Nahm data usually consists of two elements.
The first is the bulk Nahm data, which are four Hermitian $k\times k$ matrices $T_{\mu}(s)$ where the $k\in\mathbb{Z}_{+}$ is the instanton charge of calorons and $s\in[-\mu_0,\mu_0]$ is a dual space coordinate.
Here $\mu_0$ determines the fundamental interval of the dual space and the circumference of $S^1$ is given by $\pi/\mu_0$.
The boundary Nahm data is a $k$-row vector $W$ of quaternion entries.
For the massless case, \textit{i.e.}, the case of the trivial holonomy, the bulk Nahm and the boundary Nahm data satisfy, respectively,
\begin{align}
		&\frac{d}{d s}T_{i}-\frac{i}{2}\varepsilon_{ijm}\left[ T_{j},\ T_{m} \right]-i\left[ T_{4},\ T_{i} \right] = 0,	\\
		&T_{i}(-\mu_0)-T_{i}(\mu_0) = \frac{1}{2}\text{tr}_{2}\left( \sigma_{i}W^{\dagger}W \right),\label{eq:matching_comdition_Nahm_data}
\end{align} 
where $\sigma_{i}$'s are the Pauli matrices, indecies $i,j,m=1,2,3$ and the trace is taken over quaternions.

The caloron gauge fields in the ordinary, or the configuration, space are obtained from the Nahm data through the zero modes of the Weyl equations with impurities at the boundaries.
The solutions of the Weyl equations are zero modes with two elements: the bulk zero modes are $2k$-vectors $\mathbf{u}_{l}$ and the boundary zero modes are $2$-vectors $\mathbf{v}_{l}$,
where $l=1,2$ indicates the two basis of the solution space of the Weyl equations,
	\begin{align}
		&\left( \mathbf{1}_{2k}\frac{d}{ds} - i( T_{\mu}(s)+x_{\mu}\mathbf{1}_{k} )\otimes e_{\mu} \right)\ \mathbf{u}_{l} = 0,	\label{eq:Weyl_bulk_1} \\
		&\Delta\mathbf{u}_{l} := \mathbf{u}_{l}(-\mu_0) - \mathbf{u}_{l}(\mu_0) = i W^{\dagger}\mathbf{v}_{l}.	\label{eq:Weyl_boundary_1}
	\end{align}
where $e_{\mu} = (-i\sigma_{i},\mathbf{1}_{2})$ are the basis of the quaternion.
These are called the bulk Weyl equation and the boundary Weyl equation, respectively.
The next step is to find two independent pair of the zero modes $(\mathbf{u}_{1}, \mathbf{v}_{1}),\ (\mathbf{u}_{2}, \mathbf{v}_{2})$, normalized as
	\begin{equation}
		\int_{I} \mathbf{u}_{a}^{\dagger}\mathbf{u}_{b} ds + \mathbf{v}_{a}^{\dagger}\mathbf{v}_{b} = \delta_{ab},
	\end{equation}
where $a,b=1,2$ and $I=[-\mu_0,\mu_0]$.
From these zero modes, we obtain the gauge connection of calorons as
	\begin{equation}
		\bigl( A_{\mu}(x) \bigr)_{ab} = \int_{I} \mathbf{u}_{a}^{\dagger}\partial_{\mu}\mathbf{u}_{b} ds + \mathbf{v}_{a}^{\dagger}\partial_{\mu}\mathbf{v}_{b}.	\label{eq:gauge_from_zero_mode}
	\end{equation}

	In the next section, we shall describe how to numerically solve the Weyl equations.

\section{Numerical Nahm transform}
In this section, we recast the scheme developed in
 \cite{Muranaka_Nakamula_Sawado_Toda_2011}, which is specialized to the case of  instanton charge $k=2$,
 in order to carry out the numerical Nahm transform for the calorons of higher instanton charge $k\geq3$.
According to \cite{Muranaka_Nakamula_Sawado_Toda_2011}, we begin with finding the solutions to the bulk Weyl equations (\ref{eq:Weyl_bulk_1}), 
followed by solving the boundary Weyl equations (\ref{eq:Weyl_boundary_1}) with use of a degree of freedom of the linear combinations of the bulk solutions.

\subsection{The bulk Weyl equations}
The bulk Weyl equations for the charge $k$ calorons are the system of linear ordinary differential equations of rank $2k$, 
and they can be numerically solved by standard procedure such as Runge-Kutta method, etc..

First, we find out  $\alpha$ independent solutions of the Weyl equations by numerically integrating from one endpoint to the direction of another and also from the other to the first one.
Thus we obtain $2\alpha$ numerical solutions $\hat{\mathbf{u}}_{i}$ totally.
Here $\alpha$, the number of the independent solutions to the bulk Weyl equations from one endpoint,
 is also equal to the number of shooting parameters on each endpoints ($s=\pm\mu_0$).
We then fabricate two independent solutions $\mathbf{u}_{l}$ for the Weyl equations by  taking linear combinations of these $2\alpha$ solutions,
		\begin{equation}
			\mathbf{u}_{l} = \left( \hat{\mathbf{u}}_1~\hat{\mathbf{u}}_2~\dots~\hat{\mathbf{u}}_{2\alpha} \right)\cdot \boldsymbol{\omega}_{l},	\label{eq:bulk_Weyl_connection_1}
		\end{equation}
where $\boldsymbol{\omega}_{l}$ are $2\alpha$-column vectors, $i.e.\ \boldsymbol{\omega}_{l} := {}^{t}(\omega^{1}_{\ l}\ \omega^{2}_{\ l}\ \dots \ \omega^{2\alpha}_{\ l}),$ to be fixed as follows.
For the monopole construction \cite{Houghton_Sutcliffe_1996}, we impose that these solutions match at the center of the interval.
For the caloron, on the other hand, we determine them by the requirement that the bulk solutions are consistent with the boundary Weyl equations \eqref{eq:Weyl_boundary_1}.

\subsection{The boundary Weyl equations}
The boundary Weyl equations are given by
		\begin{equation}
			\Delta \mathbf{u}_{l} = iW^{\dagger}\mathbf{v}_{l}. \iff 
					\begin{pmatrix}
						\Delta u^{1}_{\ l} \\ \Delta u^{2}_{\ l} \\ \vdots \\ \Delta u^{2k}_{\ l}
					\end{pmatrix}
					= i
					\begin{pmatrix}
						W_{1}^{\dagger} \\ W_{2}^{\dagger} \\ \vdots \\ W_{k}^{\dagger}
					\end{pmatrix}
					\begin{pmatrix}
						v^{1}_{\ l} \\ v^{2}_{\ l}
					\end{pmatrix}.	\label{eq:Weyl_boundary_2}
		\end{equation}
where $W = \left( W_{1}\ W_{2}\ \dots W_{k} \right),\;W_i\in\mathbb{H}$ are the boundary Nahm data of $k$-caloron.

Although the coupled equations \eqref{eq:Weyl_boundary_2} look like an \textit{over-determined} system for two unknowns $v^{1}_{\ l}$ and $v^{2}_{\ l}$,
this is not the case because the left-hand-side are still unknown at the present stage.
Our goal is to obtain the coefficients of the linear combination $\boldsymbol{\omega}_{l}$ in \eqref{eq:bulk_Weyl_connection_1} 
and the boundary zero modes $\mathbf{v}_{l}$ simultaneously from $\hat{\mathbf{u}}_{1}, \hat{\mathbf{u}}_{2},\dots, \hat{\mathbf{u}}_{2\alpha}$.
We can reduce the problem in the form of simple linear algebra as will be shown below. 
The caloron Nahm data certainly induce $\alpha \geq k$, thus without loss of generality, we choose $\alpha=k$.
		
If we solve just the upper two rows in \eqref{eq:Weyl_boundary_2}, we can obtain the boundary zero modes $\mathbf{v}_{l}$ 
		\begin{align}
			&\begin{pmatrix}
				v^{1}_{\ l} \\ v^{2}_{\ l}
			\end{pmatrix}
			= -iW^{\dagger\ -1}_{1}	\begin{pmatrix}
							\Delta u^{1}_{\ l} \\ \Delta u^{2}_{\ l}
						\end{pmatrix}	\notag \\[3pt]
			&= \frac{-i}{\mathcal{W}^{1}_{\ 1}\mathcal{W}^{2}_{\ 2}-\mathcal{W}^{1}_{\ 2}\mathcal{W}^{2}_{\ 1}}
				\begin{pmatrix}
					\mathcal{W}^{2}_{\ 2}\Delta u^{1}_{\ l}-\mathcal{W}^{1}_{\ 2}\Delta u^{2}_{\ l}	\\
					-\mathcal{W}^{2}_{\ 1}\Delta u^{1}_{\ l}+\mathcal{W}^{1}_{\ 1}\Delta u^{2}_{\ l}
				\end{pmatrix},	\label{eq:boundary_data_to_bulk_3}
		\end{align}
where $\mathcal{W}^{i}_{\ j}$ are the components of $W^{\dagger}$ as a $2k\times2$ matrix and we used the fact that, in general, $\det W_1\neq0$.
Substituting \eqref{eq:boundary_data_to_bulk_3} into the $2(k-1)$ remaining equations in \eqref{eq:Weyl_boundary_2} (\textit{i.e.},  \eqref{eq:Weyl_boundary_2} without upper two rows), we obtain $2(k-1)$ conditions.
We solve these restraint conditions by using the degree of freedom of the linear combination of the bulk solutions. 
		
From the definition \eqref{eq:bulk_Weyl_connection_1}, $\Delta\mathbf{u}_{l}$ is expanded as
		\begin{equation}
			\Delta \mathbf{u}_{l} := (\Delta \hat{\mathbf{u}}_{1}\ \Delta \hat{\mathbf{u}}_{2}\ \dots \ \Delta \hat{\mathbf{u}}_{2\alpha})\cdot \boldsymbol{\omega}_{l}.	\label{eq:gap_bulk_solution}
		\end{equation}
where $\Delta\hat{\mathbf{u}}_{i}:=\hat{\mathbf{u}}_{i}(-\mu_0)-\hat{\mathbf{u}}_{i}(\mu_0)$.
From \eqref{eq:boundary_data_to_bulk_3} and \eqref{eq:gap_bulk_solution}, the remaining equations of \eqref{eq:Weyl_boundary_2} turn out  to be, 
		\begin{widetext}
		 \begin{align}
			&\hspace{20pt}
			\begin{pmatrix}
				\Delta u^{3}_{\ l} \\ \vdots \\ \Delta u^{2k}_{\ l}
			\end{pmatrix}
			= i
			\begin{pmatrix}
				W_{2}^{\dagger} \\ \vdots \\ W_{k}^{\dagger}
			\end{pmatrix}
			\frac{-i}{\mathcal{W}^{1}_{\ 1}\mathcal{W}^{2}_{\ 2}-\mathcal{W}^{1}_{\ 2}\mathcal{W}^{2}_{\ 1}}
				\begin{pmatrix}
					\mathcal{W}^{2}_{\ 2}\Delta u^{1}_{\ l}-\mathcal{W}^{1}_{\ 2}\Delta u^{2}_{\ l}	\\
					-\mathcal{W}^{2}_{\ 1}\Delta u^{1}_{\ l}+\mathcal{W}^{1}_{\ 1}\Delta u^{2}_{\ l}
				\end{pmatrix},	\notag \\[3pt]
			&\iff \left[
				\begin{pmatrix}
				W_{2}^{\dagger} \\ \vdots \\ W_{k}^{\dagger}
				\end{pmatrix}
				\begin{pmatrix}
					\mathcal{W}^{2}_{\ 2}\Delta \hat{u}^{1}_{\ j}-\mathcal{W}^{1}_{\ 2}\Delta \hat{u}^{2}_{\ j}	\\
					-\mathcal{W}^{2}_{\ 1}\Delta \hat{u}^{1}_{\ j}+\mathcal{W}^{1}_{\ 1}\Delta \hat{u}^{2}_{\ j}
				\end{pmatrix}
				-\det(W^{\dagger}_{1})
				\begin{pmatrix}
					\Delta \hat{u}^{3}_{\ j} \\ \vdots \\ \Delta \hat{u}^{2}_{\ j}
				\end{pmatrix} \right] \omega_{l}^{\ j} = 0,	\hspace{20pt}j=1,2,\dots,2\alpha,	\notag \\[3pt] 
			&\iff	\underbrace{ \left[ \mathcal{W}^{i}_{\ m}\gamma^{mn}_{\hspace{10pt}op}\mathcal{W}^{o}_{\ n}\Delta \hat{u}^{p}_{\ j} - \det(W^{\dagger}_{1}) \Delta \hat{u}^{i}_{\ j} \right] }_{=: A^{i}_{\ j}}\omega_{l}^{\ j} = 0, \label{eq:mulanaka_method_2_2}
					\hspace{10pt}i=3,4,\dots,2k,\; \mbox{and}\hspace{7pt} m,n,o,p=1,2,
		 \end{align}
		\end{widetext}
		where $\omega_{l}^{\ j}$ are the components of $\boldsymbol{\omega}_{l}$ and $\gamma^{mn}_{\hspace{10pt}op}$ are tensor defined as,
		\begin{equation*}
			\gamma^{mn}_{\hspace{10pt}op} :=
				\begin{cases}
					0	& m=n	\\
					1	& m\neq n
				\end{cases}\ \times \
				\begin{cases}
					+1	& n=o	\\
					-1	& n\neq o
				\end{cases}\ \times \
				\begin{cases}
					0	& o=p	\\
					1	& o\neq p
				\end{cases}.
		\end{equation*}
In \eqref{eq:mulanaka_method_2_2}, we define $(2k-2)\times2\alpha$ matrix $A$ with components $A^{i}_{\ j}$.
We can regard \eqref{eq:mulanaka_method_2_2}  as  linear homogeneous equation for  $\boldsymbol{\omega}_{l}$, 
which determine the basis of the solutions to the bulk Weyl equations.
In order to find  a basis for the $2$-dimensional kernel of $A$, we perform standard Gauss elimination.
Having obtained the independent $\boldsymbol{\omega}_{l}$, we take the linear combination \eqref{eq:bulk_Weyl_connection_1}, and find the components of $\mathbf{u}_{l}$ automatically satisfy the boundary Weyl equations \eqref{eq:Weyl_boundary_2}.
The solutions of the boundary Weyl equations $\mathbf{v}_{l}$ have already been given by \eqref{eq:boundary_data_to_bulk_3}.
		
Finally we notice that the numerical Nahm transform for the calorons in this process is known to cause a singularity lines for the action density \cite{Muranaka_Nakamula_Sawado_Toda_2011}.
The origin of the singularity lines are  the gap of the zero modes, and we can eliminate the singularity lines using the appropriate procedure,
which will be given in a forthcoming paper.

\section{Nahm data of the $C_3$-symmetric 3-caloron}
The bulk Nahm data of the  3-caloron  with $C_3$-symmetry are given by \cite{Nakamula_Sawado_2012},
\begin{subequations} \label{eq:Nahm_data_C3c}
	 \begin{align}
		T_{1} &= \frac{1}{2}
			\begin{pmatrix}
				0	& f_{+}-if_{-}	& f_{0}	\\
				f_{+}+if_{-}	& 0	& f_{+}-if_{-}	\\
				f_{0}	& f_{+}+if_{-}	& 0
			\end{pmatrix},	\\
		T_{2} &= \frac{1}{2}
			\begin{pmatrix}
				f_{0}	& -f_{+}-if_{-}	& 0	\\
				-f_{+}+if_{-}	& 0	& f_{+}+if_{-}	\\
				0	& f_{+}-if_{-}	& -f_{0}
			\end{pmatrix},	\\
		T_{3} &= \frac{1}{4}
			\begin{pmatrix}
				-p_{2}	& 0	& i(p_{0}-p_{1})	\\
				0	& 2p_{2}	& 0	\\
				-i(p_{0}-p_{1})	& 0	& -p_{2}
			\end{pmatrix},
	 \end{align} 
\end{subequations}
where $f_{\pm}:= \left( f_{1}\pm f_{2} \right)/2$, 
and  $f_{j},p_{j}\,(j=0,1,2)$ are defined by Jacobi theta functions $\vartheta_{\nu}(s,q)\;(0<q<1)$ \cite{Whittaker Watson}, 
	 \begin{align}
	\nonumber
		&f_{0}(s) := iC\frac{ \sqrt{\vartheta_{\nu}(s_{1})\vartheta_{\nu}(s_{2})} }{ \vartheta_{\nu}(s_{0}) },	
		f_{1}(s) := iC\frac{ \sqrt{\vartheta_{\nu}(s_{2})\vartheta_{\nu}(s_{0})} }{ \vartheta_{\nu}(s_{1}) },	\\
		&\hspace{2cm}f_{2}(s) := iC\frac{ \sqrt{\vartheta_{\nu}(s_{0})\vartheta_{\nu}(s_{1})} }{ \vartheta_{\nu}(s_{2}) },
	 \end{align}
	and
	 \begin{align}
	\nonumber 
		&p_{0}(s) := \frac{d}{ds}\log\frac{\vartheta_{\nu}(s_{0})}{\vartheta_{\nu}(s_{2})},	
		p_{1}(s) := \frac{d}{ds}\log\frac{\vartheta_{\nu}(s_{1})}{\vartheta_{\nu}(s_{0})}, \\
		&\hspace{2cm}p_{2}(s) := \frac{d}{ds}\log\frac{\vartheta_{\nu}(s_{2})}{\vartheta_{\nu}(s_{1})}.
	 \end{align}
In the expression, we omit the explicit $q$ dependence for simplicity.
For the hermiticity of the bulk data, the index of the theta functions have to be $\nu=0$ or $3$, and the constant is chosen such that $C:=\vartheta_{1}'(0)/\vartheta_{1}(1/3) \in i\mathbb{R}$.
In addition, we fix $s_{j}:=\pm s+j/3$ for the periodicity $f_{j+3}=f_{j}$ and $p_{j+3}=p_{j}$ due to $\vartheta_{\nu}(s_{j+3})=\vartheta_{\nu}(s_{j})$.
We also find that the bulk data enjoy the reality conditions $T_{i}(-s)={}^{t}T(s)$.
Note that the $C_3$-symmetry is not manifest in this basis of the bulk data \cite{Nakamula_Sawado_2012}.

Next we consider the boundary data $W$.
The boundary data in the charge $3$ case are given by $3$-row vector of quatenion entries
	\begin{equation}
		W=\left( \lambda,\rho,\chi \right).
	\end{equation}
From the matching conditions \eqref{eq:matching_comdition_Nahm_data} and using the bulk data \eqref{eq:Nahm_data_C3c}, we obtain
the $C_3$-symmetric boundary data,
	 \begin{align}
	\nonumber 
		&\lambda = i\lambda_{1}\left( \sigma_{1}+\sigma_{2} \right),	
		\rho = -\frac{2}{\lambda_{1}}g(\mu_0) \mathbf{1}_{2},	\\
		&\hspace{1cm}\chi = -i\lambda_{1}\left( \sigma_{1}-\sigma_{2} \right),
	 \end{align}
where  $g(\mu_0):= -f_{-}(\mu_0)$ and $\lambda_1$ is a real parameter constrained by $2\lambda_1^{2} = h(\mu_0)>0$
with the definition
	\begin{align}
		h(\mu_0) := -\frac{1}{2}\left( p_{0}(\mu_0)-p_{1}(\mu_0) \right).\label{the constraint}
	\end{align}
Note that the constraint gives a restriction on the bulk data
	\begin{equation}
		p_{0}(\mu_0)-p_{1}(\mu_0) <0.	\label{eq:C3c_Nahm_data_condition_1}
	\end{equation}
Taking into account the constraint, we find that the $C_3$-symmetric 3-caloron has one free parameter $q$.

For the  calorons having monopole limit, \textit{e.g.}, \cite{Muranaka_Nakamula_Sawado_Toda_2011, Ward_2004}, 
the bulk Nahm data are regular on the fundamental period, say, $s\in(-\ell/2,\ell/2)$, and have simple poles at $s=\pm \ell/2$ where the residues span an irreducible representation of $su(2)$.
Hence, $\mu_0$ takes values in the range $(0, \ell/2)$. 
In other words, it has upper bound which corresponds to the monopole limit.
For  the  $C_3$-symmetric 3-calorons considering here, however, the Nahm data is regular in $s\in\mathbb{R}$, and consequently, $\mu_0$ will take values from zero to infinity.
	
In order to satisfy the condition \eqref{eq:C3c_Nahm_data_condition_1}, the acceptable theta function depends on the value of $\mu_0$.
Introducing the term ``sector", which is open interval $M_n:=(n/2,(n+1)/2)$ for each $n\in\mathbb{Z}_{\geq0}$, we find 
	\begin{equation}
		\vartheta_{\nu} :=
			\begin{cases}
				\vartheta_{3},	& \mbox{for}\; \mu_0\in M_{2n},	\\
				\vartheta_{0},	& \mbox{for}\; \mu_0\in M_{2n+1}.
			\end{cases}	\label{eq:C3c_Nahm_data_condition_2}
	\end{equation}
Now the sign of the argument of the theta function is arbitrary chosen as $s_{j}=\pm s+j/3$.
From the profile of the bulk Nahm data, we find $h(n/2)=g(n/2)=0,\ n\in\mathbb{Z}_{\geq0}$.
This leads all of the components of the boundary data $W$ is zero, and the boundary zero mode is not well-defined at the boundary of the sectors, $\mu_0=n/2$.
Consequently,  when $\mu_0=n/2$, the $C_3$-symmetric 3-caloron Nahm data is not well-defined.

\section{The action density}
In order to get further insight of the solutions, it is surely worthwhile to visualize the action density, which of course is gauge invariant and positive definite.
From the ASD conditions $\ast F_{\mu\nu}=\pm F_{\mu\nu}$, where $\ast$ is the Hodge dual on $\mathbb{R}^{4}$, the action density $S$ of calorons can 
be written as
	\begin{equation}
		S(x^{i},t) := -\frac{1}{2}\text{tr}_{2}F_{\mu\nu}^{2} = -2\text{tr}_{2}\left( F_{12}^{2}+F_{23}^{2}+F_{31}^{2} \right).
	\end{equation}
Here the field strength tensor is defined as $F_{\mu\nu}:=\partial_{\mu}A_{\nu}-\partial_{\nu}A_{\mu} + \left[ A_{\mu},A_{\nu} \right]$.
Thanks to the ASD conditions, we do not have to calculate the  $t = x^{4}$ derivative in the field strength so that we can regard $t$ as a parameter in the calculation.

In the following, we fix $t=x^{4}=0$.
We employ 151 grid points for the dual space, and $71\times 71\times 71$ lattice points for the configuration space, which are sufficient for the numerical convergence.
We perform the visualization of the action density mainly by Mathematica \cite{Mathematica}.

As mentioned in the previous section, the Nahm data of  $C_3$-symmetric 3-caloron has a notable, special feature;
the period of the dual space $\mu_0$ takes values in $(0,\ \infty)$ except for the isolated points $n/2,\;n\in\mathbb{Z}_{\geq0}$.
Thus it will be interesting when we plot the action density for the cases of large period.
Since our main concern is to clarify the behavior for the change of $\mu_0$, we plot the action densities for fixed $q$.
	
	\subsection{The first two sectors: $M_0,M_1$}
\begin{figure*}[htbp]
		 \begin{center}
			\includegraphics[width=15cm]{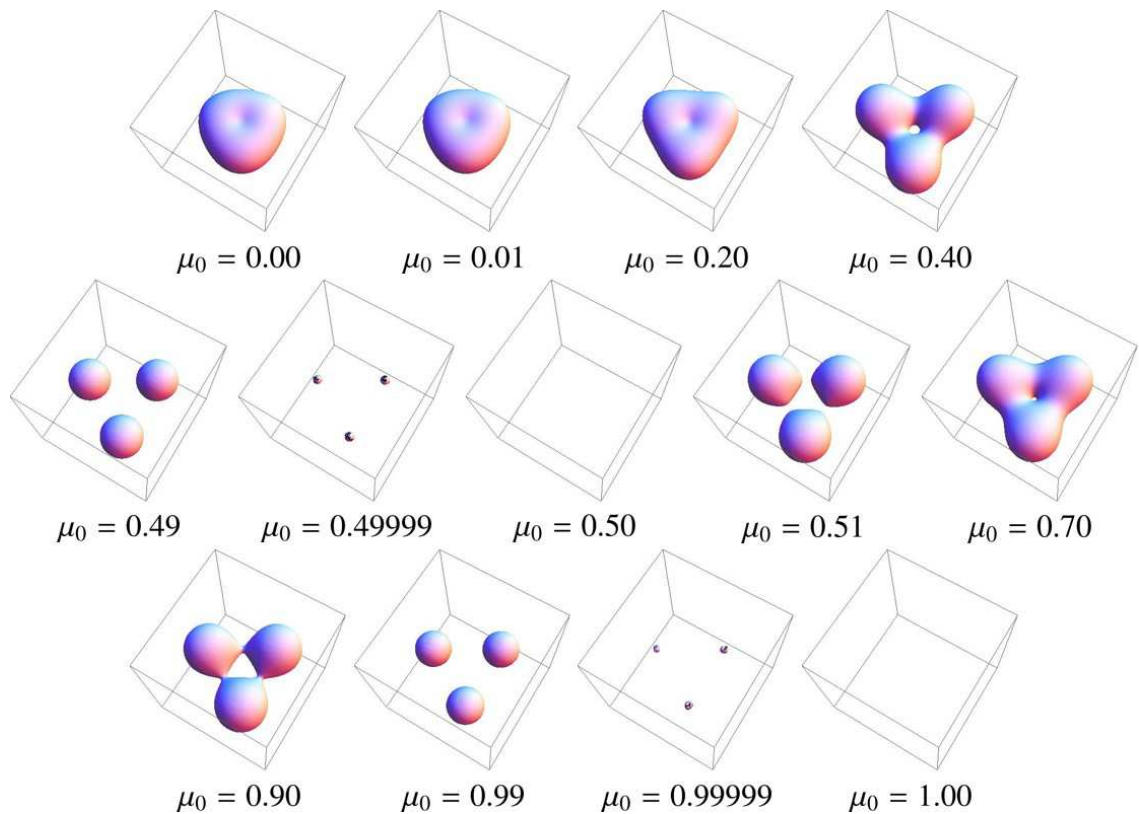}
			\caption{Action isosurface plot of the cyclic 3-caloron with $q=0.1,\ S(x^i,t)=7.0,\ t=0.0$ and $x^i\in(-2.0,2.0)$ respectively.
	Note that the plots $\mu_0=0.50$ and $1.00$ are actually the cases of  $\mu_0=0.4999999999999999$ and $\mu_0=0.999999999999999$, respectively.			
	Because the boundary Nahm data is not well-defined at exactly $\mu_0=0.50$ and $1.00$.  
	Also,  the plot $\mu_0=0.00$  is the case $\mu_0=0.0000000001$.
				}	\label{figure:ac+_q=0.1}
		 \end{center}
		\end{figure*}

In Fig.\ref{figure:ac+_q=0.1}, we present the action isosurface plots for values of $\mu_0\in M_0=(0.0,\; 0.5),\;M_1=(0.5,\; 1.0)$.
For the sector $M_0$, we use  $\vartheta_3$, while for the sector $M_1$ we use $\vartheta_0$, according to (\ref{eq:C3c_Nahm_data_condition_2}).
One can manifestly see that all the densities exhibit $C_{3}$-symmetry.
As was already pointed out, the Nahm data are not well-defined at $\mu_0=0.5,$ and $1.0$.
We find that the action isosurface gradually shrinks as $\mu_0$ increases and finally disappears at $\mu_0=0.5$.
After passing $\mu_0=0.5$, it appears again and grows, and repeatedly it reduces and finally vanishes at $\mu_0=1.0$.		
Note that in the figure of $\mu_0=0.0$, we use the value $\mu_0=0.0000000001$ instead of $\mu_0$ being exactly zero, because we cannot perform the numerical calculation for the case.

	\subsection{The ``exterior" sectors: $M_n~(n\geq 2)$}
\begin{figure*}[htbp]
		 \begin{center}
			\includegraphics[width=10cm]{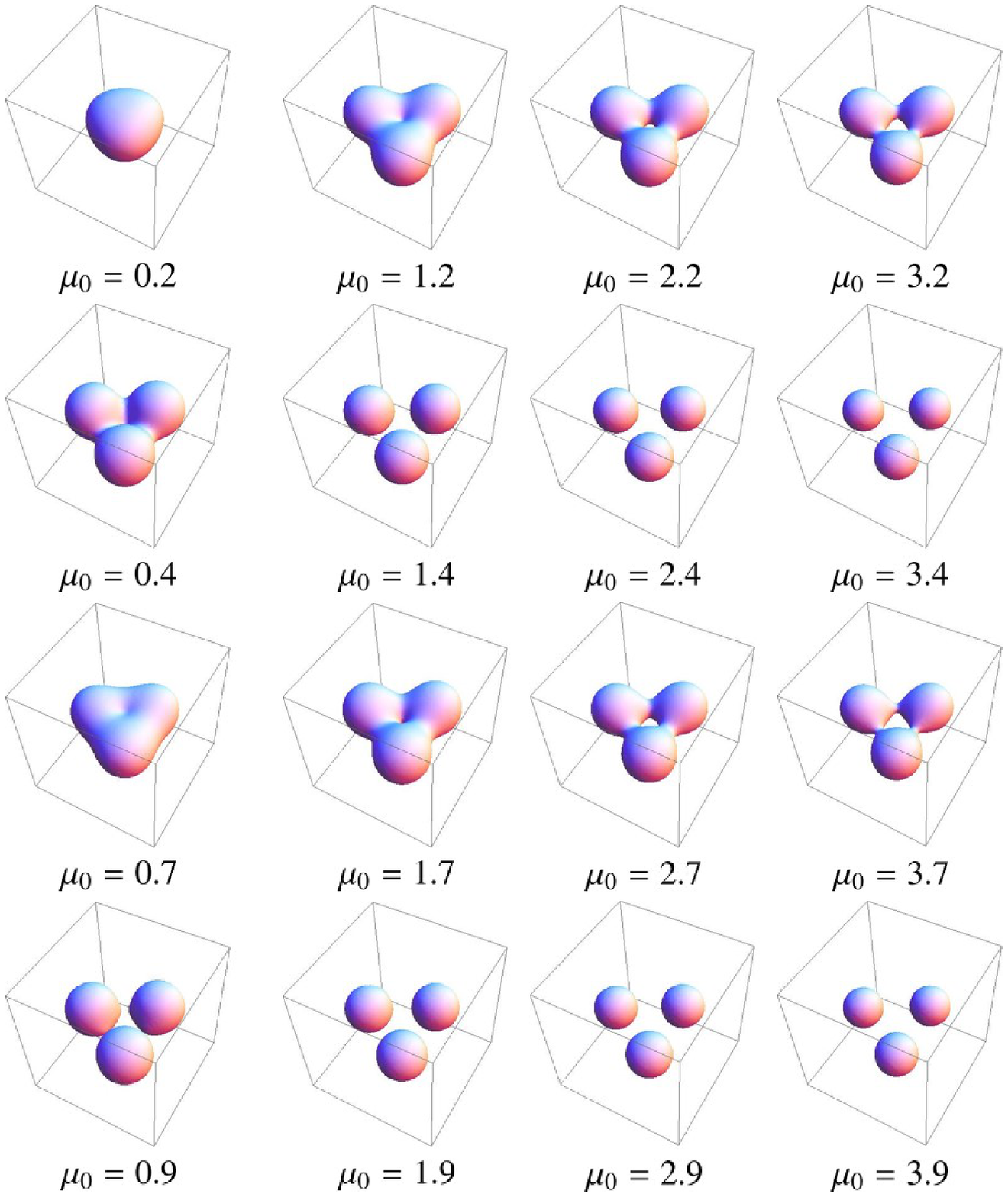}
			\caption{Action isosurface plot  with $q=0.3,\ S(x^i,t)=7.0\ t=0.0$ and $x^i\in(-2.0,2.0)$, respectively.}	
			\label{figure:acu+_q=0.3}
		 \end{center}
		 \begin{center}
			\includegraphics[width=10cm]{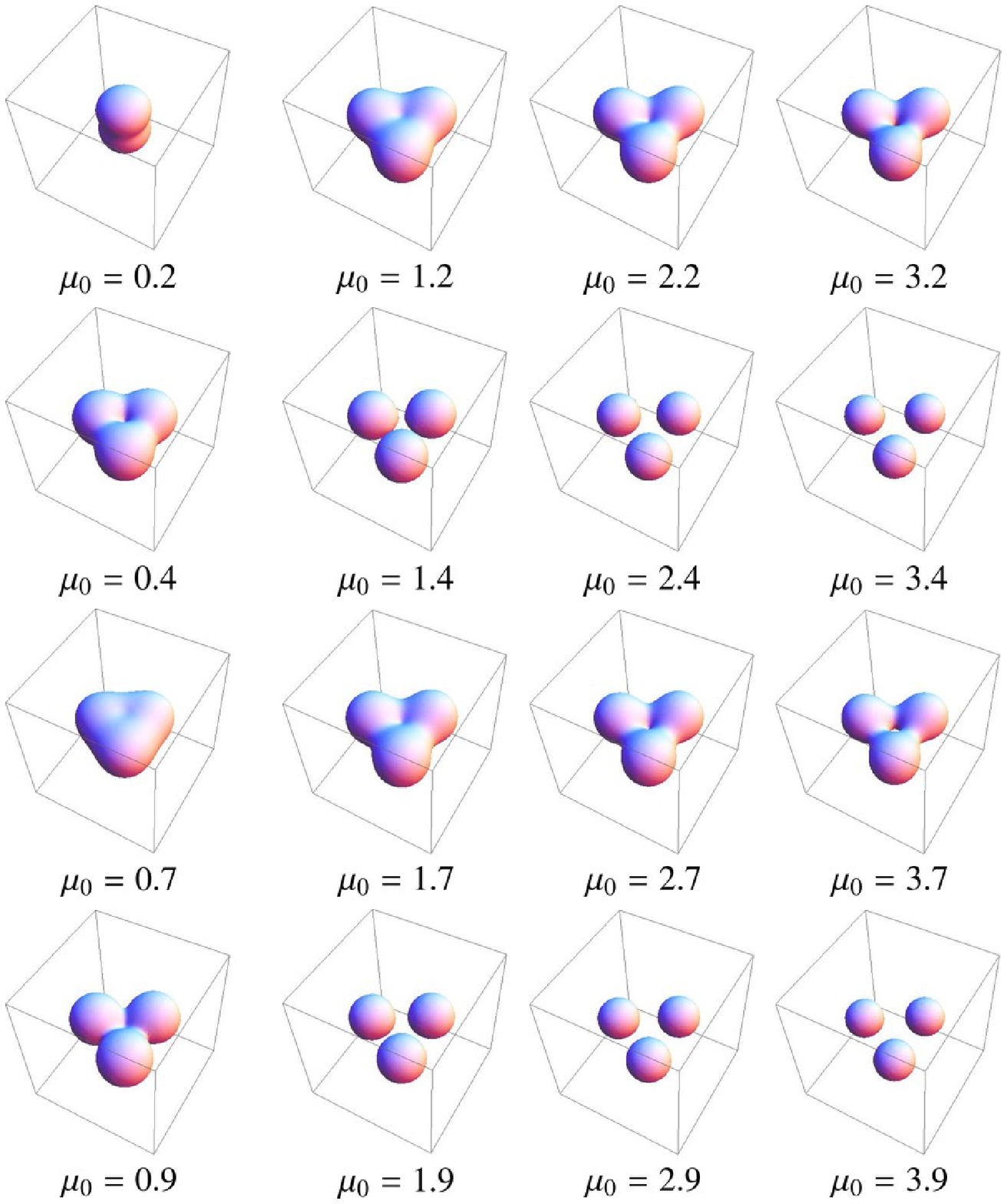}
			\caption{Action isosurface plot  with $q=0.5,\ S(x^i,t)=7.0\ t=0.0$ and $x^i\in(-2.0,2.0)$, respectively.}	
			\label{figure:acu+_q=0.5}
		 \end{center}
		\end{figure*}

In Figs.\ref{figure:acu+_q=0.3} and \ref{figure:acu+_q=0.5}, we present the isosurface plots of the action density for larger values of $\mu_0$
with fixed $q$ as $0.3$ and $0.5$.
From these plots, we see a periodic behavior of the isosurfaces with fixed $\mu_0(\notin M_0)$ and $\mu_0+n,\;(n=1,2,\cdots)$, 
namely, the similarity of the plots in the horizontal rows.
However, the calculation for $\mu_0>1.0$ shows that the action densities are reducing gradually as $n$ increases.
We expect the action density tends to vanish for $\mu_0\to\infty$, which corresponds to the case $\mathbb{R}^3\times S^1\to\mathbb{R}^3$ in the configuration space.
The detail of this point will be given in elsewhere.

\section{conclusion}
In this paper, we have constructed a general scheme for numerical Nahm transform of calorons with arbitrary instanton charge.
As a first application, we have performed the Nahm transform of the $C_3$-symmetric 3-calorons and found the quasi-periodic behavior of the action density 
as  $\mu_0$ varies.
The bulk Nahm data are given by Jacobi theta functions $\vartheta_3$ and $\vartheta_0$ sector by sector.

Although $\mu_0$, the inverse of the circumference of $S^1$, can take almost all positive values, the half-integer values 
$\mu_0=n/2, \;n\in\mathbb{Z}_{\geq0}$ are excluded due to the ill-definition of the Nahm data.
In particular, the case  $\mu_0=0.0$, corresponding to the instanton on $\mathbb{R}^4$, is also eliminated from the consideration.
At this point, we need complete understanding to the relation with the exact ADHM data of $C_3$-symmetric instanton obtained in \cite{Nakamula_Sawado_2012}.
We expect to report the subject in next articles.

{\bf Acknowledgement}
The authors would like to thank D.Muranaka for fruitful discussions and comments.

\end{document}